\pdfoutput=1 
\documentclass[a4paper, 11pt]{article}
\usepackage[dvipsnames]{xcolor}
\usepackage{jheppub}
\usepackage{physics}
\usepackage{braket}
\usepackage{scalerel}
\usepackage{blindtext} 
\usepackage{epsfig, cancel}
\usepackage{latexsym}
\usepackage{natbib, comment}
\usepackage[all]{xy}
\makeatletter
\gdef\@fpheader{}
\makeatother
\usepackage[T1]{fontenc}
\usepackage{amsmath,amssymb,amsthm,amsfonts,tikz,graphicx,hyperref,lmodern}
\usepackage{stmaryrd}
\usepackage[utf8]{inputenc}
\usepackage{bm,mathrsfs}
\hypersetup{ linktoc=all,
    colorlinks, linkcolor=blue,
    citecolor=magenta, urlcolor=red
}

\def\mathbi#1{\textbf{\em #1}}
\newcommand{\n}{\mathbi{n}}
\newcommand{\m}{\mathbi{m}}

\newcommand{\inbox}[1]{\tcbset{fonttitle=\scriptsize} \tcboxmath[colback=white,colframe=black!70]{#1}}

\newcommand{\be}{\begin{equation}}
\newcommand{\ee}{\end{equation}}
\newcommand{\bea}{\begin{eqnarray}}
\newcommand{\eea}{\end{eqnarray}}

\newcommand{\PP}{\text{P}\!\!\!\text{P}}
\newcommand{\PPhi}{\text{P}\!\!\Phi}
\newcommand{\PhiPhi}{\Phi\!\!\!\Phi}

\definecolor{rosy}{RGB}{230,235,252}
\definecolor{myframetitle}{RGB}{90,89,170}
\definecolor{myblocktitle}{RGB}{140,185,249}
\definecolor{mytitle}{RGB}{10,80,26}
\definecolor{darkgreen}{RGB}{27,130,45}
\definecolor{darkblue}{rgb}{0,0,0.3}
\definecolor{darkred}{rgb}{0.7,0,0}
\definecolor{light gray}{RGB}{220,220,220}
\definecolor{dark purple}{RGB}{108,0,217}
\definecolor{pink}{RGB}{190,20,100}
\definecolor{orang}{RGB}{193,63,0}
\definecolor{green}{RGB}{11,98,17}
\definecolor{darkpink}{RGB}{153,0,76}
\definecolor{bluegreen}{RGB}{0,102,102}
\definecolor{greenlagan}{RGB}{0,102,0}
\definecolor{redgreen}{RGB}{102,102,0}
\definecolor{Redgreen}{RGB}{153,76,0}
\definecolor{vividviolet}{rgb}{0.62, 0.0, 1.0}
\definecolor{amaranth}{rgb}{0.9, 0.17, 0.31}
\definecolor{palatinateblue}{rgb}{0.15, 0.23, 0.89}
\definecolor{brightpink}{rgb}{1.0, 0.0, 0.5}
\definecolor{cornflowerblue}{rgb}{0.39, 0.58, 0.93}
\definecolor{deepcarminepink}{rgb}{0.94, 0.19, 0.22}
\definecolor{radicalred}{rgb}{1.0, 0.21, 0.37}
\definecolor{darkmagenta}{rgb}{0.67, 0, 0.67}

\makeatletter \@addtoreset{equation}{section}

\newcommand{\cV}{\mathcal{V}}
\newcommand{\bcM}{\boldsymbol{\mathcal{M}}}
\newcommand{\cW}{\mathcal{W}}
\newcommand{\uQ}{\underline{Q}}

\usepackage[most]{tcolorbox}
\tcbset{highlight math style={left=02mm,right=02mm,top=02mm,bottom=02mm}}
\usepackage{empheq}

\begin{document}

\title{{\LARGE{\centerline{Null Strings Gauged, Reloaded and Quantized, II:}}} 
\centerline{\Large{Null Strings on Flat Backgorund Have Discrete Spectrum}} }   

\author[a]{Ida M. Rasulian}
\author[a]{, M.M. Sheikh-Jabbari}
\author[b]{, H. Yavartanoo}
\affiliation{$^a$ School of Physics, Institute for Research in Fundamental Sciences (IPM), P.O.Box 19395-5531, Tehran, Iran}
\affiliation{$^b$ Beijing Institute of Mathematical Sciences and Applications (BIMSA), Huairou District, Beijing 101408, P. R. China}
\emailAdd{idarasulian@ipm.ir, jabbari@theory.ipm.ac.ir, yavar@bimsa.cn}

\abstract{We continue our study of the quantization of null strings, strings with Carrollian worldsheets described by an action with two-dimensional ($2d$) diffeomorphisms, Weyl and Carroll-Weyl scaling as gauge symmetry. We perform  light cone gauge quantization of the null strings on flat space background within Schr\"odinger representation, initiated in \cite{Rasulian:2026jvg}. We show, within the regular separated Schr\"odinger sector, that bosonic null strings on a flat background have no critical dimensions, have discrete, nonnegative mass-squared levels and are hence free of tachyonic modes.}

\maketitle
\flushbottom

\section{Introduction}\label{sec:Intro}

Null string theory (NST) is a described by a two-dimensional $(2d)$ Carrollian worldsheet theory in which all the geometric redundancies of the $2d$ worldsheet are gauged; these geometric redundancies are $2d$ diffeomorphisms, Weyl scaling and Carroll-Weyl (CW) scaling \cite{Sheikh-Jabbari:2026cnj, Sheikh-Jabbari:2026vqh, Sheikh-Jabbari:2026tpf, Sheikh-Jabbari:2026kwr}. The NST is different than the ILST theory \cite{Isberg:1993av} which does not exhibit the essential CW gauge-invariance. As in the standard string theory which has $2d$ Lorentzian (or Euclidean) worldsheet and described by Polyakov action \cite{Polyakov:1981rd, Polchinski:1998rq}, NST can be formulated on different background target spaces \cite{Sheikh-Jabbari:2026rcr, Sheikh-Jabbari:2026nez, Duary:2026zar, Lindstrom:2026tua}, nonetheless in this work we focus on the NST on $D$ dimensional flat Minkowski target space. 

Having a first principle definition of classical NST, the next step is to carry out quantization of the theory. There are different quantization procedures for Polyakov strings, like canonical or path integral quantizations \cite{Polchinski:1998rq, Green:1987sp}. These quantization schemes may be performed in a covariant way  introducing ghosts or in lightcone gauge (LCG) by solving all the constraints associated with gauge symmetries and complete gauge-fixings \cite{Polchinski:1998rq,Green:1987sp}. In the NST case, besides the standard $bc$ ghost system, we have a third ghost associated with the CW gauge symmetry, hence dealing with $bcs$ ghost system \cite{Rasulian:2026jvg, Duary:2026lmk, Chen:2026klv, Chen:2026cau}. In this work, however, we perform quantization in the Schr\"odinger wavefunctional scheme, outlined and implemented in \cite{Rasulian:2026jvg}.

We quantize free NST on flat background in the LCG and solve the Schr\"odinger equation with lightcone Hamiltonian for Schr\"odinger wavefunctionals for null strings. This method is well-suited the NST specially because it allows us to bypass potential theoretical subtleties of BRST symmetries in non-Lorentzian theories and allows us to discuss a suitable vacuum (ground-state) and work out the spectrum of the theory. 

Despite dealing with a ``tensionless'' string \cite{Schild:1976vq, Isberg:1993av}, the reduced kernel equation has discrete positive eigenvalues in a regular class specified below. We extend the quadratic construction of \cite{Rasulian:2026jvg} to arbitrary functional order and prove that for a generic, regular eigenfunctions of the unprojected representative equation are polynomial on each ordered region.  Its translation into a complete physical spectrum requires, in addition, control of projection onto the nonzero modes, the separation conditions, and the physical pairing, which we carefully specify. The {quantization scheme in the specified sector selects no critical dimension. Moreover, the mass-squared of  null strings is positive-definite and hence there is no tachyonic mode in the bosonic null strings.}

This work is organized as follows. Section~\ref{sec:REVIEW} reviews basics of null string theory in the lightcone gauge and its Schr\"odinger quantization scheme. Section~\ref{sec:Higher-order-wavefunctionals} derives the kernel eigenvalue equation for generic Schr\"odinger wavefunctionals and shows that the separation conditions
allow its projector to be removed.  Section~\ref{sec:Polynomial-solutions} constructs its exact polynomial solutions. Section~\ref{sec:Global-kernels} treats global matching, a weighted spectral realization, and polynomial exhaustion under explicit regularity assumptions. Section~\ref{sec:discussion} is devoted to discussions and outlook. The appendices contain technical details of derivations.

\section{Null strings review}\label{sec:REVIEW}

In this section we quickly review classical null string theory on flat space in the lightcone gauge (LCG).  This is a concise review of earlier work in \cite{Sheikh-Jabbari:2026cnj, Sheikh-Jabbari:2026vqh, Sheikh-Jabbari:2026tpf}. We then review Schr\"odinger quantization scheme and apply it to the null string theory in the LCG; this part is a quick review of \cite{Rasulian:2026jvg}. 

\subsection{Classical null strings in the LCG}\label{sec:REVIEW-LCG-fixing}

The NST on $D$ dimensional flat Minkowski background is governed by the action which realizes $2d$ diffeomorphisms, Weyl and CW scalings as gauge symmetry \cite{Sheikh-Jabbari:2026vqh, Sheikh-Jabbari:2026tpf, Sheikh-Jabbari:2026kwr}:
\begin{equation}\label{NST-Action}
S = \frac{\kappa}{2} \int \text{d}^2\sigma\, \mathcal V^a \mathcal V^b D_aX^\mu D_bX_\mu ,
\end{equation}
where $\sigma^a=(\tau, \sigma)$  denote the worldsheet coordinates. We use the normalized spatial integration measure
\begin{equation}
\int \mathrm{d}^2\sigma
:=\int d\tau\int_0^{2\pi}\frac{d\sigma}{2\pi}.
\end{equation}
The momentum field is defined relative to this measure, so that
in temporal gauge $P_\mu=\kappa\dot X_\mu$. ${\cal V}^a$  encodes the kernel vector of the $2d$ Carrollian and $X^\mu$ are the string embedding coordinates and 
\begin{equation}
    D_a X^\mu= \partial_a X^\mu + \cW_a X^\mu
\end{equation}
where $\cW_a$ is the CW gauge field. This action is invariant under diffeomorphisms+Weyl+CW scalings, because $\cV^a, X^\mu$ are invariant under ordinary Weyl scaling and under the other two generated by $\eta=(\xi^a, \chi)$ where $\xi^a=\xi^a(\tau,\sigma), \chi=\chi(\tau, \sigma)$ \cite{Sheikh-Jabbari:2026vqh}
\begin{equation}\label{eta-transform}
\begin{split}
\delta_\eta X^\mu&= \xi^a\partial_a X^\mu + \chi X^\mu\\
  \delta_\eta {\cal V}^a &= \xi^b\partial_b{\cal V}^a - {\cal V}^b\partial_b\xi^a  +(\frac12\partial_b\xi^b-\chi) {\cal V}^a\\
    \delta_\eta {\cal W}_a &=\xi^b\partial_b{\cal W}_a + {\cal W}_b\partial_a\xi^b  -\partial_a\chi \,.
\end{split}\end{equation}
With the above $\delta_\chi(D_aX^\mu)=\chi D_aX^\mu$. To analyze the theory we need to fix the two symmetries. This may be conveniently done in the $\cV^a\cW_a=0$ and  $\mathcal V^a = (1,0)$ gauges, upon which the gauge-fixed action reduces to, 
\begin{equation}\label{NST-gauge-fixed-Action}
S = \frac{\kappa}{2} \int \text{d}^2\sigma\, \dot X^\mu\dot X_\mu ,\qquad \dot X:=\partial_\tau X, 
\end{equation}
The momentum field is defined relative to this normalized measure,
so that $P_\mu=\kappa\dot X_\mu$.
This is the ILST action in the temporal gauge \cite{Isberg:1992ia, Isberg:1993av}. This gauge-fixing leaves us with three constraints \cite{Sheikh-Jabbari:2026cnj}
\begin{align}
\mathcal C_1 = P^2 \approx 0 , \qquad \mathcal C_2 = P\cdot X' \approx 0 , \qquad \mathcal C_3 = P\cdot X \approx 0
,\qquad X':=\partial_\sigma X 
\end{align}
and the residual gauge transformations, 
\begin{equation}\label{residual-transform}
\delta_\eta X^\mu= \left(h+(f'-2\chi)\tau\right) \dot X^\mu + f X'^\mu+ \chi X^\mu
\end{equation}
where $h, f, \chi$ are functions of $\sigma$ (and not $\tau)$.

The EoM in the temporal gauge and its most general solutions are,
\begin{equation}\label{GeneralSolution}
\ddot X^\mu =0 \quad \Longrightarrow \quad X^\mu(\tau,\sigma) = Q^\mu(\sigma) + \frac{1}{\kappa} P^\mu(\sigma) \tau .
\end{equation}
With this decomposition the constraints will be
\begin{equation}\label{Constraints-onshell} 
 \texttt{C}_1:=P^2(\sigma) = 0 , \qquad 
\texttt{C}_2:=P(\sigma) \cdot Q'(\sigma) = 0 , \qquad 
\texttt{C}_3:= P(\sigma) \cdot Q(\sigma) = 0 . 
\end{equation}

\paragraph{Null Strings in the Light-cone gauge.} The above constraints can be explicitly solved adopting the $\chi$-fixed light-cone gauge (LCG) \cite{Sheikh-Jabbari:2026tpf, Rasulian:2026jvg}
\begin{equation}\label{chi-LCG-fixing}
\begin{split}
X^+ = \frac{1}{\sqrt{2}}\left(X^0+ X^{D-1}\right)&=\frac{{p}{}^+}{\kappa}\tau,\qquad p^+>0 \\
Q^1(\sigma)= q_1&, \qquad q_1\neq 0
\end{split}
\end{equation}
and labeling the target-space-time  coordinates as 
\begin{equation}
X^\mu = \left(X^+,X^-,X^1,X^I\right),
\qquad
I=2,\ldots,D-2.
\end{equation}
The $\texttt{C}_1=0$, $\texttt{C}_3=0$ constraints can be solved as
\begin{equation}
P^- = \frac{(P_1)^2 + P^I P^I}{2{p}^+}, \qquad Q^- = \frac{P_1 q_1 + P^I Q^I}{{p}^+}.
\label{PMinusSolution}
\end{equation}
The remaining constraint $\texttt{C}_2=0$ leads to
\begin{equation}\label{C2-NSLCG}
P^{1\prime} q_1 + P^{I\prime} Q^I = 0,
\end{equation}
which we solve as
\begin{equation}\label{P1Solution}
P^1(\sigma) =  \hat{p}_1- \frac{1}{q_1} \int_{0}^{\sigma} d\bar\sigma\, P^{I\prime}(\bar\sigma) Q^I(\bar\sigma),\qquad \hat{p}_1=P_1(\sigma=0)\,.
\end{equation}
From this we find the global integrability condition
\begin{equation}\label{eq:global_integrability}
\int_0^{2\pi} d\sigma\, P^{1\prime}(\sigma) = 0 \implies \int_0^{2\pi} d\sigma\, P^{I\prime}(\sigma)Q^I(\sigma) = 0. 
\end{equation}

Expanding in Fourier modes
\begin{align}\label{eq:XMode_rev}
P^I(\sigma) = p^I + \sum_{n\neq0} P_n^I e^{in\sigma}, \qquad 
Q^I(\sigma) = q^I + \sum_{n\neq 0} Q_n^I e^{in\sigma}, 
\end{align}
we find
\begin{equation}\label{eq:P1ModeExact}
\begin{split}
    P^1(\sigma) = & p_1 + \sum_{n\neq0} P_n^1 e^{in\sigma},\\
p_1=\hat{p}_1+\frac{q^I}{q_1}\sum_{n\neq 0} P^I_n {+\frac{1}{q_1}\sum_{n\neq0}\Phi_n} &,\qquad P_n^1 = -\frac{q^I}{q_1} P_n^I - \frac{1}{q_1} \Phi_n,  ,
\end{split}
\end{equation}
where 
\begin{equation}\label{Phin-def}
    \Phi_n := \sum_{r\neq0, n} \frac{r}{n} P_r^I Q_{n-r}^I, \qquad n\neq0\,.
\end{equation}
The global integrability condition \eqref{eq:global_integrability} isolates a single global constraint mode:
\begin{equation}\label{eq:G0Review}
G_0 := \sum_{n\neq0} nP_n^I Q_{-n}^I = 0\,.
\end{equation}

The light cone Hamiltonian is

\begin{equation}\label{eq:Hamiltonian_Full_Rigorous}
H_{\mathrm{LC}}
:=\frac{1}{2\pi}\int_0^{2\pi}d\sigma\,P^-(\sigma)
=\frac{1}{2p^+}
\left[(p_1)^2+p^Ip^I+{\cal M}^2\right].
\end{equation}
where we have
\begin{equation}\label{M2-classical}
{\cal M}^2= {\cal G}_{IJ}\ {\text{P}\!\!\!\text{P}}^{IJ} + \frac{1}{(q_1)^2} \left(2q^I \text{P}\!\!\Phi_I +  \Phi\!\!\!\Phi\right)
\end{equation}
with 
\begin{equation}
{\cal G}^{IJ} := \delta^{IJ}+ \frac{q^I q^J}{(q_1)^2}, \qquad {\PP}^{IJ}:=\sum_{n\neq0} P_{-n}^I P_n^J,\qquad \PPhi_I:=\sum_{n\neq0}  P_{-n}^I \Phi_n,\qquad   \PhiPhi:=\sum_{n\neq0} \Phi_{-n} \Phi_n .
\end{equation}

\paragraph{Mass function.}
The light-cone Hamiltonian has the dispersion-relation form
\begin{equation}\label{mass-function}
{\mathcal M}^{2}=2p^+H_{\mathrm{LC}}-(p_1)^2-p^Ip^I.
\end{equation}
This motivates identifying ${\mathcal M}^{2}$ as the
mass-squared function  in the reduced light-cone description. Noting that $p_I, p_1, H_{\mathrm{LC}}, p^+$ are components of the momentum associated with center of mass of the null string (averages over $\sigma$), ${\mathcal M}^{2}$ may be viewed as the mass of the center of mass of the null string. Interestingly, except for the ground state, configurations with vanishing $P_n^I$, which has vanishing ${\mathcal M}^{2}$, generic null string configuration has non-zero ${\mathcal M}^{2}$. This means that, despite the fact that null strings is a congruence of null geodesics (recall $\texttt{C}_1$) \cite{Sheikh-Jabbari:2026kwr}, the center of mass is not following a null geodesic. {We also note that $P^1(\sigma)$ has a nonlocal expression in terms of $P^I(\sigma), Q^I(\sigma)$. This nonlocality has no counterpart in the ILST theory or standard tensile string theory in the light-cone gauge and arises once one solves $\texttt{C}_3$ to express $P^1(\sigma)$ in terms of the transverse modes and is manifested through $\Phi_n$ \cite{Rasulian:2026jvg}.  }

In the next section, we promote ${\mathcal M}^{2}$ to the mass operator whose eigenvalues define the conditional spectral problem studied below.

\paragraph{LCG reduced (physical) phase space.} The LCG-fixed solution space description contains the transverse fields $Q^I(\sigma),P_I(\sigma)$ subject to level-matching condition $G_0=0$, at sectors with fixed given  $p^+, q_1,p_1$. With the normalized spatial measure, their canonical brackets are
\begin{equation}\label{canonical-PB}
\{q^I,p^J\}=\delta^{IJ},\qquad
\{Q_n^I,P_m^J\}=\delta^{IJ}\delta_{n+m,0},
\qquad n,m\neq0.
\end{equation}
These brackets describe the reduced variables before imposing
$G_0=0$ and identifying configurations related by the rigid
$\sigma$-translations generated by $G_0$. See Appendix~\ref{app:qnsA-reduction} for a summary of LCG phase space. 

{At fixed nonzero $q_1$, its magnitude sets an effective length scale for the null string theory and its spectrum. Introduce dimensionless variables through the substitutions 
\begin{equation}\label{q1-removal}
    p_1 \to \frac{p_1}{q_1}, \qquad
    p^I\to \frac{p^I}{q_1},\qquad
    q^I\to q_1q^I,\qquad
   Q_n^I\to q_1Q_n^I,\qquad
 P_n^I\to \frac{P_n^I}{q_1}.
\end{equation} 
Here the expressions on the right replace the original variables, and the rescaled variables retain their original symbols. At fixed $q_1$, the transverse canonical brackets are preserved,  and ${\cal G}^{IJ}=\delta^{IJ}+q^I q^J$. In terms of the rescaled variables, $q^I,Q_n^I$, their canonical conjugates and ${\cal M}^2$ are all dimensionless and $q_1$ dependence in ${\cal M}^2$ as well as in $H_{\mathrm{LC}}$, only appears in a prefactor $1/q_1^2$. Hereafter, we will adopt this parameterization where $q_1$, which is a label on solution sectors of NST in flat space, appears as its effective length scale.
In the spectral problem below, these data are held fixed, while
the transverse variables $q^I,Q_n^I$ are quantized.
This defines a conditional quantum problem; quantization and
dynamical evolution of the pair $(q_1,p_1)$ are outside the scope
of the present analysis.

\subsection{Schr\"odinger quantization scheme for null strings in the LCG}\label{sec:REVIEW-LCG-Quantization}

We quantize the transverse canonical variables in the
Schr\"odinger representation, with $p^+,q_1,p_1$ held fixed
as specified above. We use the dimensionless variables
of \eqref{q1-removal} and associate a wavefunctional
\begin{equation}
\Psi = \Psi\!\left[ q^I, \{Q_n^I\}_{n\neq0};\ {p}^+,  p_1; x^+ \right],
\label{eq:WavefunctionalDef}
\end{equation}
which satisfies Schr\"odinger equation and the null-string level matching condition:
\begin{equation}
H_{\text{LC}} \Psi = i\ \partial_{x^+} \Psi,\qquad \widehat{G}_0 \Psi=0
\label{Schrod-Eq-x+}
\end{equation}
In the $Q$ representation,
\begin{equation}
p^I = -i \frac{\partial}{\partial q^I}, \qquad
P_n^I = -i \frac{\partial}{\partial Q_{-n}^I}.
\label{eq:SchrodingerDifferentialOperators}
\end{equation}
and in terms of the rescaled variables of \eqref{q1-removal}, the $H_{\text{LC}}$ and $\widehat{G}_0$ take  the form \cite{Rasulian:2026jvg}
\begin{equation}\label{HLC-X-basis}
H_{\text{LC}} = \frac{1}{2{p}^+ q_1^2} \left[ (p_1)^2 - \frac{\partial^2}{\partial q^I\partial q^I} + \boldsymbol{{\cal M}}^2\right],\qquad \widehat{G}_0=i\sum_{n\neq 0} n \ Q_{n}^I\ \frac{\delta}{\delta Q_n^I}
\end{equation}
with
\begin{equation}\label{M2-quantum}
\boldsymbol{{\cal M}^2}= -{\cal G}^{IJ}\sum_{n\neq 0}
\frac{\delta^2}{\delta Q_{-n}^I\delta Q_n^J}
+2q^I\sum_{\substack{m,n\neq0\\m+n\neq0}}\frac{m}{n}Q_{n+m}^J
\frac{\delta^2}{\delta Q_m^J\delta Q_n^I}+\boldsymbol{\PhiPhi},
\qquad {\cal G}^{IJ}=\delta^{IJ}+q^Iq^J.
\end{equation}
We define the last term by composition of the reconstructed operators,
\begin{equation}\label{eq:b0-product-prescription}
\boldsymbol{\PhiPhi}:=\sum_{n\neq0}\widehat\Phi_{-n}\widehat\Phi_n,
\qquad
\widehat\Phi_n=i\sum_{\substack{r\neq 0, -n}}
\frac{r}{n}Q_{n+r}^I\frac{\partial}{\partial Q_r^I}.
\end{equation}

We note that  $\Phi_n$ has no ordering ambiguity, as $\hat\Phi_n \cdot= i\sum_{r\neq 0, -n}  \frac{r}{n}  Q_{n+r}^I\frac{\delta}{\delta Q^I_{r}} \ \cdot =i\sum_{r\neq 0, -n}  \frac{r}{n}  \frac{\delta}{\delta Q^I_{r}} (Q_{n+r}^I\ \cdot)$. The $\PP$ term has no ordering ambiguity because $P^I_n$ commute among themselves. Similarly, the $\PPhi$ term does not have ordering ambiguity, because $[P_{-n}^I,\Phi_n]=0$.  Recall that, 
\begin{equation}\label{eq:b0-opposite-commutator}
 \begin{aligned}
[\widehat\Phi_n,\widehat\Phi_{-n}]=
-\frac2n\sum_{r\neq0}rQ_r^I\frac{\partial}{\partial {Q^I_r}}
=\frac{2i}{n}\widehat G_0.
\end{aligned}
\end{equation}
Thus, on level-matched states, the commutator vanishes and  the $\PhiPhi$ term is also free of ordering ambiguity. 
Written in the ordering where all the derivatives w.r.t $Q^I_n$ are pushed to the right, we obtain
\begin{equation}\label{PhiPhi-quantum}
\boldsymbol{\PhiPhi}=
\sum_{\substack{n,r,s\neq0\\n\neq s,\ n\neq-r}}
\frac{rs}{n^2}Q_{n+r}^IQ_{-n+s}^J
\frac{\delta^2}{\delta Q_r^I\delta Q_s^J}
+\frac{\pi^2}{3}\sum_{r\neq0}r^2Q_r^I\frac{\delta}{\delta Q_r^I}.
\end{equation}
In the notation of \cite{Rasulian:2026jvg}, with the normalization
of the first-derivative term in \eqref{PhiPhi-quantum}, this is
$\gamma=1$, hence $b=(1-\gamma)\pi^2/3=0$. In the above, we have considered $\bcM^2$ as an operator made out of $P^I $ and $\Phi$, and the latter is treated as a composite operator. 

\paragraph{Vacuum state.} The vacuum in this framework  satisfies 
\begin{equation}
p^I \Psi_0 = 0, \qquad P_n^I \Psi_0 = 0, \qquad \forall n \neq 0.
\label{eq:VacuumCondition}
\end{equation}
From this we find that $\Psi_0$ is a constant. In the terminology of the ``three vacua'' for tensionless strings \cite{Bagchi:2020fpr}, the above corresponds to ``induced vacuum''.  The other two vacua do not yield a consistent null string theory \cite{Chen:2026cau, Chen:2026klv, Duary:2026lmk} and there is no  critical dimension for null strings on Minkowski flat space. One should, however, note that 

the constant functional is a formal vacuum: its normalizability requires a measure and a domain that have not yet been specified. The absence of a dimension-dependent obstruction in the reduced differential equations does not by itself establish closure of all quantum target-space symmetry generators.

\subsection{Quadratic wavefunctionals}
\label{sec:quadratic-wavefunctionals}
Within an expansion in homogeneous powers of the nonzero modes,
the vacuum state $\Psi_0$ \eqref{eq:VacuumCondition} has degree zero.
A linear state in the nonzero-mode sector is excluded by level matching. The first
nontrivial homogeneous nonzero-mode sector is therefore quadratic:
\begin{equation}
\Psi_2= e^{ik_I q^I}\psi_2,\qquad \psi_2=\sum_{r\neq0}K^r_{IJ}Q_{-r}^IQ_r^J\,\Psi_0.
\label{eq:physicalquadratic}
\end{equation}

We take the coefficients $K^r_{IJ}$ to be independent of $q^I$
and require the mass eigenvalue equation to hold for every
$q^I$ with the same eigenvalue. The expansion in homogeneous
powers of $Q^I_n$ modes specifies the class of states
considered here.

Using
\begin{equation}
Q_r^I=\frac1{2\pi}\int_0^{2\pi}d\sigma\,e^{-ir\sigma}Q^I(\sigma),
\qquad K_{IJ}(\theta)=\sum_{r\neq0}e^{ir\theta}K^r_{IJ},
\qquad \int_0^{2\pi}K_{IJ}(\theta)d\theta=0,
\end{equation}
the state may be written in the bilocal form
\begin{equation}
\psi_2=\frac1{(2\pi)^2}\int_0^{2\pi}d\sigma\int_0^{2\pi}d\widetilde\sigma\,
Q^I(\sigma)Q^J(\widetilde\sigma)K_{IJ}(\sigma-\widetilde\sigma)\,\Psi_0.
\label{eq:BilocalKernel}
\end{equation}
The zero-average condition chooses a representative of the 
kernel. With $\uQ^I(\sigma):=Q^I(\sigma)-q^I$, the same functional
can be written using $\uQ$ and an arbitrary representative: a constant
addition to $K$ is invisible. Only the part satisfying
$K_{IJ}(\theta)=K_{JI}(-\theta)$ contributes.

For the zero-average representative one finds
\begin{equation}
\bcM^2\psi_2=-2{\cal G}^{IJ}K_{IJ}(0)\Psi_0+
\boldsymbol{\PhiPhi}\psi_2.
\end{equation}
{
The $\PPhi$ term vanishes by level matching in the quadratic
sector. A pure quadratic mass eigenstate therefore requires
${\cal G}^{IJ}K_{IJ}(0)=0$ for every $q^I$.
Together with $K_{IJ}(0)=K_{JI}(0)$, this gives $K_{IJ}(0)=0$.
With these conditions,
}
\begin{equation}
\boldsymbol{\PhiPhi}\psi_2=
\frac{2}{(2\pi)^2}\int d\sigma\,d\widetilde\sigma\,
\uQ^I(\sigma)\uQ^J(\widetilde\sigma)
\Upsilon(\sigma-\widetilde\sigma)K_{IJ}''(\sigma-\widetilde\sigma)\,\Psi_0,
\end{equation}
where 
\begin{equation}\label{eq:betaFourier}
\Upsilon(\theta)=\sum_{n\neq0}\frac{e^{in\theta}}{n^2}-\frac{\pi^2}{3}
=\frac{\theta^2}{2}-\pi|\theta|,\qquad -2\pi\leq\theta\leq2\pi.
\end{equation}
It is continued periodically, and
\begin{equation}
\partial_\theta^2\Upsilon=1-2\pi\sum_{n\in\mathbb Z}\delta(\theta+2\pi n),
\qquad \Upsilon(\theta+2\pi)=\Upsilon(\theta).
\end{equation}
Here integration by parts is understood periodically, including
the coincidence contribution in $\Upsilon''$, which implies
\begin{equation}
 \int_0^{2\pi}\Upsilon(\theta)K_{IJ}''(\theta)\,d\theta
 =
 \int_0^{2\pi}K_{IJ}(\theta)\,d\theta-2\pi K_{IJ}(0)=0.
\end{equation}
Thus the mass eigenvalue equation yields the \textbf{master equation} for the kernel $K$:
\begin{equation}\label{eq:BbetaEq}
\inbox{\Upsilon(\theta)K_{IJ}''(\theta)
=\frac12\lambda(\lambda+1)K_{IJ}(\theta),\qquad
\mu^2=\lambda(\lambda+1).}
\end{equation}
{We note that at the ``coincidence point''  $\sigma_1=\sigma_2$ or equivalently $\theta=0, 2\pi$, $\Upsilon(0)=\Upsilon(2\pi)=0$. Hence, for $\lambda\neq 0$, smoothness of $K$ at the coincidence point implies that $K(0)=K(2\pi)=0$. This is in accord with the $\PP$ requirement discussed above. (Note that derivatives of $K$ remain bounded at the coincidence point, while they can have finite jumps.)}

Using the fact that $\Upsilon$ is a $2\pi$-periodic even function, one may introduce the $t$ variable, 
\begin{equation}\label{eq:xCoordinate}
t=\frac{\theta-\pi}{\pi},\qquad -1\leq t\leq1,
\end{equation}
in terms of which the master equation~\eqref{eq:BbetaEq} becomes
\begin{equation}\label{eq:traceODEt}
(1-t^2)K''(t)+\lambda(\lambda+1)K(t)=0.
\end{equation}
The transformation
\begin{equation}\label{eq:BtoF}
K(t)=\sqrt{1-t^2}f(t)
\end{equation}
yields the associated Legendre equation \cite{Gradshteyn-Ryzhik},
\begin{equation}\label{eq:assocLegendreDerived}
(1-t^2)f''-2tf'+
\left[\lambda(\lambda+1)-\frac1{1-t^2}\right]f=0.
\end{equation}
 For solutions which vanish and are regular at both endpoints $t=-1,1$ the solutions are (associated) Legendre polynomials.
The corresponding family is
\begin{equation}\label{GIJ-Sym-full-range}
K^r_{IJ}(t)=\sqrt{1-t^2}\ K^{r0}_{IJ}\ P^1_{\lambda_r}(t)
=-K^{r0}_{IJ}\ (1-t^2)P'_{\lambda_r}(t),\qquad -1\leq t\leq1.
\end{equation}
Here $K^{r0}_{IJ}$ is a constant tensor and $r=s,a$ denotes its
symmetric or antisymmetric part, including the trace in the symmetric
part. The kernels vanish at coincidence, $\theta=0\equiv2\pi$,
which corresponds to $t=\pm1$. The zero-average condition excludes
$\lambda=1$, since it is proportional to $-(1-t^2)$ with integral
$-4/3$ over $[-1,1]$. All degrees $\lambda\geq2$ have zero average.
Exchange symmetry therefore gives
\begin{equation}\label{eq:b0-quadratic-levels}
\lambda_a=2k,\quad \mu_a^2=2k(2k+1),\quad k\geq1;
\qquad
\lambda_s=2k-1,\quad \mu_s^2=2k(2k-1),\quad k\geq2.
\end{equation}
We extend these kernels periodically and continuously.
Their first derivatives may jump at coincidence, where
$\Upsilon$ vanishes, so the equation remains valid
distributionally.

\section{Beyond quadratic wavefunctionals: kernel equations and physical constraints}
\label{sec:Higher-order-wavefunctionals}
The quadratic construction describes states whose dependence on the string
profile is encoded by a kernel of two {insertion positions.} 
To go beyond this sector,
we allow products of three or more transverse coordinates $Q^I(\sigma)$. Their coefficient
kernels describe how the state depends on the profile at several points
along the string. The spectral problem then becomes a differential equation
in their relative positions, supplemented by the conditions inherited from
the mass eigenvalue equation. This section formulates
this problem; polynomial solutions of the kernel equation and their global
matching will be studied separately. 

{
As in the quadratic case, we consider functionals of the
nonzero modes,
\begin{equation}\label{eq:qns3-separable}
\Psi= e^{ik_I q^I}\psi[\uQ], \qquad 
 \uQ^I(\sigma)=Q^I(\sigma)-q^I
 =\sum_{n\neq0}Q_n^I e^{in\sigma},
 \qquad
 \int_0^{2\pi}d\sigma\,\uQ^I(\sigma)=0.
\end{equation}
For a real string profile, $Q_{-n}^I=(Q_n^I)^*$.
}

\subsection{Higher-order wavefunctionals and their kernels}
\label{subsec:qns3-wavefunctionals}

We reviewed the quadratic wavefunctionals in the previous section.
In this and the following sections, we study the homogeneous polynomial degree $N$ ($N>2$) wavefunctionals 
\begin{equation}
\psi_{_N}[\uQ] =\sum_{I_1,\ldots,I_N} \int\prod_{a=1}^N d\sigma_a\, K_{I_1\ldots I_N}(\sigma_1,\ldots \sigma_N) \prod_{a=1}^N Q^{I_a}(\sigma_a). \label{eq:qns3-functional}
\end{equation}
We work with  representative kernels with vanishing average in each argument,
\begin{equation}\label{K-int0}
\int_0^{2\pi} d\sigma_a\, K_{I_1\ldots I_N}(\sigma_1,\ldots \sigma_N)=0, \qquad a=1,\ldots,N,
\end{equation}
for every choice of the remaining arguments and indices.
Consequently, replacing each $Q^I(\sigma)$ in
\eqref{eq:qns3-functional} by the zero-mode subtracted part $\uQ^I$ leaves the wavefunctional unchanged.
The expression therefore depends only on $\uQ$. {\color{blue}

}

The kernel arguments are $N$ integration positions on the spatial
circle of the closed string, as illustrated in
Fig.~\ref{fig:qns-kernel-circle}.
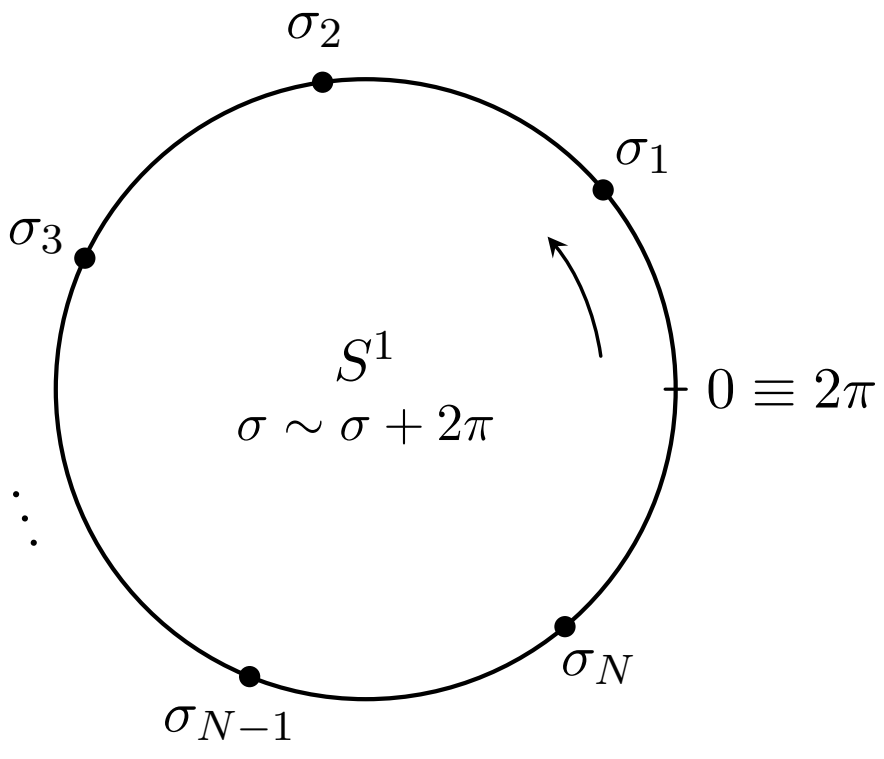
\begin{figure}[h]
 \centering
 \begin{tikzpicture}[line cap=round,line join=round,font=\normalsize]
  \draw[line width=0.8pt] (0,0) circle[radius=2.05cm];

  \draw[line width=0.6pt] (1.98,0) -- (2.12,0);
  \node[anchor=west,inner sep=3pt] at (2.14,0) {$0\equiv2\pi$};

  \foreach \angle/\slot in {40/1,98/2,155/3,248/{N-1},310/N}{
   \fill (\angle:2.05cm) circle[radius=2pt];
   \node at (\angle:2.40cm) {$\sigma_{\slot}$};
  }
  \node[rotate=110,inner sep=0pt] at (200:2.40cm) {$\cdots$};

  \node at (0,0.22) {$S^1$};
  \node[font=\small] at (0,-0.24) {$\sigma\sim\sigma+2\pi$};
  \draw[->,>=stealth,line width=0.65pt]
   (8:1.57cm) arc[start angle=8,end angle=40,radius=1.57cm];
 \end{tikzpicture}
 \caption{Integration positions $\sigma_1,\ldots,\sigma_N$ of the kernel
 $K_{I_1\ldots I_N}(\sigma_1,\ldots,\sigma_N)$ on the spatial circle
 of the closed string at fixed worldsheet time. Each position is paired
 with a factor $\uQ^{I_a}(\sigma_a)$ in $\psi_{_N}[\uQ]$; the marked points
 are integration variables, not a discretization of the string.
 The circle represents the periodic worldsheet coordinate, not the
 target-space shape. Level matching imposes invariance under a common
 shift $\sigma_a\mapsto\sigma_a+\epsilon$ modulo $2\pi$, leaving
 $N-1$ independent relative positions.}
 \label{fig:qns-kernel-circle}
\end{figure}
Since the coordinate factors commute, only the kernel symmetric
under simultaneous exchanges of coordinate and tensor slots
contributes. We therefore choose
\begin{equation}
K_{\ldots I_a\ldots I_b\ldots} (\ldots,\sigma_a,\ldots,\sigma_b,\ldots)=K_{\ldots I_b\ldots I_a\ldots} (\ldots,\sigma_b,\ldots,\sigma_a,\ldots).
\label{eq:qns3-slot-symmetry}
\end{equation}
This is not symmetry of the tensor indices alone.
An antisymmetric tensor structure can contribute when accompanied
by the corresponding antisymmetry in the position dependence.
Closed-string periodicity is imposed in each position separately.

The remaining rigid translation of the string coordinate imposes level
matching,
\begin{equation}
 \sum_{n\ne0,I}n Q_n^I\ \frac{\partial\psi_{_N}}{\partial Q_n^I}=0\quad \Longrightarrow \quad \sum_a\partial_{\sigma_a}K=0.  \label{eq:qns3-level-matching}
\end{equation}
That is, the state depends on relative positions, not on an origin along the closed string. Equivalently, \eqref{K-int0} and \eqref{eq:qns3-level-matching} imply,
\begin{equation}
 K_{I_1\ldots I_N}(\boldsymbol\sigma)  =\sum_{\substack{r_1,\ldots,r_N\ne0\\r_1+\cdots+r_N=0}} {\cal K}_{I_1\ldots I_N}(r_1,\ldots,r_N) e^{i\sum_a r_a\sigma_a},
 \label{eq:qns3-fourier-kernel}
\end{equation}
where all $r_a$ are integers and ${\cal K}$ has the same simultaneous slot
symmetry. With the unnormalized integration measure in
\eqref{eq:qns3-functional},
\begin{equation}
 \psi_{_N}  =(2\pi)^N  \sum_{\substack{I_1,\ldots,I_N\\r_1,\ldots,r_N\ne0\\ r_1+\cdots+r_N=0}}  {\cal K}_{I_1\ldots I_N}(r_1,\ldots,r_N)\  Q_{-r_1}^{I_1}\cdots Q_{-r_N}^{I_N}. 
 \label{eq:qns3-fourier-functional}
\end{equation}
This form makes both the absence of zero-mode factors and mode-number
conservation explicit. It does not impose a reality condition on the
quantum state, which may be complex. This wavefunctional should also be supplemented with  periodicity, regularity, and other  physicality constraints. 

\subsection{Level \texorpdfstring{$N$}{N} 
homogeneous wavefunctional mass eigenstates}
\label{subsec:qns3-remaining-conditions}

We are now ready to study mass eigenvalue equations:
\begin{equation}\label{M2-eigenvalue-eq}
    \bcM^2 \psi_{_N}= \mu^2 \psi_{_N}
\end{equation}
with $\bcM^2$ given in \eqref{M2-quantum}. As in the quadratic case, the same mass eigenvalue
equation must hold for every $q^I$. Since $\psi_{_N}$ is
independent of $q^I$,  this implies
\begin{equation}
    \PP^{IJ}\psi_{_N}=0,\qquad  \PPhi_I\psi_{_N}=0\,.
\end{equation}
In terms of  Fourier modes, 
\begin{align}
 &\sum_{n\ne0}
 \frac{\partial^2\psi_{_N}}
 {\partial Q_{-n}^{I}\partial Q_n^{J}}=0,
 &&\text{for every }I,J,
 \label{eq:qns3-Q-condition}\\
 &\sum_J\sum_{\substack{n,m\ne0\\n\ne m}}
 \frac{m}{n}\,Q_{n-m}^{J}
 \frac{\partial^2\psi_{_N}}
 {\partial Q_n^{I}\partial Q_{-m}^{J}}=0,
 &&\text{for every }I.
 \label{eq:qns3-R-condition}
\end{align}

Eq.\eqref{eq:qns3-Q-condition} has a simple interpretation in terms of mode-number conservation. For $N\geq2$, it yields 
\begin{equation}
\sum_{n\ne0} {\cal K}_{IJ I_3\ldots I_N} (-n,n,r_3,\ldots,r_N)=0, \qquad r_3,\ldots,r_N\ne0,\quad r_3+\cdots+r_N=0,
\label{eq:qns3-pair-contraction}
\end{equation}
This condition holds for every choice of the remaining mode
numbers and tensor indices.
For $N=3$, the contraction would leave a single factor with mode number
zero. Such a factor is absent from the wavefunctional of the nonzero modes, so 
\eqref{eq:qns3-Q-condition} holds for every level-matched homogeneous cubic 
wavefunctional. Nonetheless, \eqref{eq:qns3-R-condition}
is not automatic: that operation can leave two nonzero, opposite modes.
Every cubic candidate must still satisfy it. At order four $N=4$, the first condition already becomes nontrivial. It includes
\begin{equation}
 \sum_{n\ne0} {\cal K}_{IJKL}(-n,n,s,-s)=0, \qquad s\ne0.
 \label{eq:qns3-quartic-Q}
\end{equation}

\paragraph{Mass eigenvalues.}

Once the $\PP, \PPhi$ conditions \eqref{eq:qns3-Q-condition} and \eqref{eq:qns3-R-condition} are satisfied, the mass eigenvalue equation \eqref{M2-eigenvalue-eq} reduces to,
\begin{equation} \label{eq:qns3-composite-eigenvalue}
 \boldsymbol{\PhiPhi}\ \psi_{_N} =\mu^2\ \psi_{_N}.
\end{equation}

The zero-average condition \eqref{K-int0}, together with
\eqref{eq:qns3-Q-condition} and \eqref{eq:qns3-R-condition},
ensures that the kernel produced by the mass operator also
has vanishing average in every argument. Hence
\eqref{eq:qns3-composite-eigenvalue} is equivalent to
\begin{equation}
 \inbox{
 \sum_{a>b}^{N}\Upsilon(\sigma_a-\sigma_b)
 \partial_{\sigma_a}\partial_{\sigma_b}K
 =-\frac{\mu^2}{2}K}
 \label{eq:qns3-unprojected-eigenvalue}
\end{equation}
with $\Upsilon$ given by \eqref{eq:betaFourier}.  For $N=2$, this equation reduces to \eqref{eq:BbetaEq}. At a coincidence point $\sigma_a=\sigma_b$, $\Upsilon$ vanishes.
Assuming the second derivatives of the kernel remain bounded
as the coincidence point is approached,
\eqref{eq:qns3-unprojected-eigenvalue} reduces to the level $N-1$ master equation.

The cubic $N=3$  sector is the first case with two independent relative positions. In an
ordered region, choose
\begin{equation}
 \theta_1={\sigma_2-\sigma_1},\qquad \theta_2={\sigma_3-\sigma_2},\qquad 
 \qquad \theta_1,\theta_2\geq 0, \qquad 0\leq \theta_1+\theta_2\leq 2\pi.
 \label{eq:qns3-gaps}
\end{equation}
The three gaps describe the successive separations around the circle, a close string (cf. Fig.~\ref{fig:qns-kernel-circle}).
Suppressing tensor indices, the representative equation becomes
\begin{equation}
 \left[  \Upsilon(\theta_1)\partial_{\theta_1}^2+  \Upsilon(\theta_2)\partial_{\theta_2}^2+ \theta_1\theta_2\;\partial_{\theta_1}\partial_{\theta_2}  \right]K(\theta_1,\theta_2)  =\frac{\mu^2}{2}K(\theta_1,\theta_2).
 \label{eq:qns3-cubic-equation}
\end{equation}
The boundaries $\theta_1=0$, $\theta_2=0$, and $\theta_1 + \theta_2=2\pi$ correspond to pairwise
coincidences, including coincidence across the periodic cut. As is explicitly seen, at the coincidence points, say $\theta_2=0$, \eqref{eq:qns3-cubic-equation} reduces to $N=2$ equation \eqref{eq:BbetaEq}. Solutions in neighboring ordered regions
must describe one periodic kernel with the exchange symmetry
\eqref{eq:qns3-slot-symmetry}.  We require continuity across coincidence points and smoothness inside each ordered region. Finite jumps in first normal derivatives are allowed when the distributional equation permits them. The polynomial-exhaustion result in Section~\ref{sec:Global-kernels} uses the stronger, explicitly stated assumption of smooth extension to the closure of each ordered region; it does not require derivatives to agree across different regions.
\section{Exact polynomial solutions}
\label{sec:Polynomial-solutions}
The master equation \eqref{eq:qns3-unprojected-eigenvalue}
is a second-order PDE for the kernel function $K$, supplemented by
periodicity and coincidence matching.  In this section we construct its polynomial solutions on an
ordered region {$\sigma_1<\sigma_2<\cdots<\sigma_N<2\pi+\sigma_1$}. Their continuation to a closed-string kernel and their
interpretation as states in the nonzero-mode sector will be considered separately.

\subsection{Decomposition of the master operator and polynomial eigenstates}
\label{subsec:qns4-chamber}

On an ordered region as depicted in Fig.~\ref{fig:qns-kernel-circle}, {$\sigma_1<\sigma_2<\cdots<\sigma_N<2\pi+\sigma_1$}, we introduce
\begin{equation}\label{eq:qns4-gaps}
 \theta_a={\sigma_{a+1}-\sigma_{a}},\quad a=1,\ldots,N-1,
 \qquad\theta_N=2\pi-\sum_{a=1}^{N-1}\theta_a,
 \qquad\theta_1,\ldots,\theta_N>0.
\end{equation}
They give the separations of successive points around the closed
string. Only $\theta_1,\ldots,\theta_{N-1}$ are independent;
$\theta_N$ measures the remaining separation across the chosen cut.
With the scaling operator
\begin{equation}
 {\cal D}=\sum_{a=1}^{N-1}\theta_a\partial_{\theta_a},
\end{equation}
the representative equation becomes
\begin{equation}\label{eq:qns4-eigenvalue}
 L_NK:=(L_0+L_{-1})K=\frac{\mu^2}{2}K,
\end{equation}
where
\begin{equation}\label{eq:qns4-operator}
 L_0=\frac12{\cal D}({\cal D}-1),\qquad
 L_{-1}=-\pi\sum_{a=1}^{N-1}\theta_a\partial_{\theta_a}^2.
\end{equation}
We have suppressed the tensor indices, as they are spectators in this calculation. For $N=3$ one
recovers \eqref{eq:qns3-cubic-equation} and a detailed derivation may be found in Appendix~\ref{app:qnsD-chamber}. 
The term $L_0$ preserves polynomial degree and
$L_{-1}$ lowers it by one. This {
allows polynomial solutions to be constructed; exhaustion of the regular class is proved in Section~\ref{sec:Global-kernels}.
} To see this, consider
\begin{equation}\label{eq:qns4-polynomial}
 K_m=h_m+h_{m-1}+\cdots+h_0,
 \qquad {\cal D}h_j=jh_j,\qquad h_m\ne0.
\end{equation}
Here $h_j$ is a homogeneous polynomial of degree $j$ in its $N-1$ arguments $\theta_a$, not necessarily a single monomial. The leading, highest degree, term fixes the eigenvalue. At lower degree the equation is
\begin{equation}\label{eq:b0-main-recursion}
 \left[\frac{m(m-1)-j(j-1)}2\right]h_j=L_{-1}h_{j+1},
 \qquad j=m-1,\ldots,0.
\end{equation}

For $m\geq2$ the bracket is positive at every lower degree, so the
completion is unique for any prescribed $h_m$. In particular,
\begin{equation}\label{eq:qns4-levels}
 L_NK_m=\frac{\mu_m^2}{2}K_m,
 \qquad \mu_m^2=m(m-1),\qquad m=2,3,\ldots.
\end{equation}
The recursion can be written in closed form:
\begin{equation}\label{eq:b0-main-completion}
 K_m=\sum_{j=0}^m
 \frac{2^{m-j}(m+j-2)!}{(m-j)!(2m-2)!}
 (L_{-1})^{m-j}h_m.
\end{equation}
Since $L_{-1}$ involves second order derivatives in all $\theta_a$, $L_{-1} h_0=0, L_{-1} h_1=0$, \eqref{eq:b0-main-recursion} implies $h_0=0$. The linear solutions at $m=0,1$ have eigenvalue zero and must be treated separately. The local multiplicity of functions $K_m$ at degree $m\geq2$ is
$\binom{m+N-2}{N-2}$. It counts homogeneous leading polynomials before
global matching, tensor-slot symmetry, and projection onto the nonzero modes. Thus, for $N=2$, there is $\binom{m}{0}=1$ case for any $m$, which is compatible with the explicit construction reviewed in section~\ref{sec:quadratic-wavefunctionals}. There, we also discussed the smoothness and matching criteria taking the $IJ$ indices into account. 

\subsection{Low-degree examples and symmetries in
\texorpdfstring{$\theta_a$}{theta} variables}
\label{subsec:qns4-examples}

It is instructive to analyze $N=3$ more closely. For generic $m$,
there are $\binom{m+3-2}{3-2}=m+1$ independent local polynomial eigenfunctions. The operator $L_3$ is invariant under cyclic permutations of $\theta_1,\theta_2,\theta_3$, subject to
$\theta_1+\theta_2+\theta_3=2\pi$. Consequently, cyclic permutations preserve each polynomial
eigenspace. A real basis can be organized into invariant solutions and pairs that mix under these permutations. 
{One may show that among $m+1$ solutions of degree $m>2$ for $N=3$,  $\lfloor \frac{m}{2}\rfloor+1$ are eigenstates of cyclic permutations and the rest rotate into each other among themselves. This follows from considering the $m+1$ chain $\theta_1^m, \theta_1^{m-1}\theta_2, ... ,\theta_2^m$ and pairing the terms that are identical up to interchanging $\theta_1$ and $\theta_2$.} Below, we illustrate this structure for $m=2, 3$.

For $m=2$, we have three independent $\theta_1(2\pi-\theta_1)$, $\theta_2(2\pi-\theta_2)$,
$\theta_1\theta_2$.  The $K_2$ solutions, which have $\mu^2=2$, can be written as
linear combinations of
\begin{equation}\label{N3-m2}
\begin{split}
 {\cal S}_1=\theta_1&\theta_2+\theta_2\theta_3+\theta_3\theta_1,\\
 {\cal M}_1=\theta_1(\theta_2-\theta_3),\quad
 {\cal M}_2 &=\theta_2(\theta_3-\theta_1),\quad
 {\cal M}_3=\theta_3(\theta_1-\theta_2)
\end{split}
\end{equation}
with ${\cal M}_1+{\cal M}_2+{\cal M}_3=0$. Under cyclic permutations, ${\cal S}_1$ is invariant while $\mathcal M_1\longmapsto\mathcal M_2 \longmapsto\mathcal M_3\longmapsto\mathcal M_1$. 

For $m=3$, with $\mu^2=6$, there are four independent {local eigenfunctions}, which may be written as linear
combinations of
\begin{equation}\label{N3-m3}
\begin{split}
 {\cal S}_1&=\theta_1\theta_2\theta_3,\\
 {\cal S}_2&=(\theta_1-\theta_2)(\theta_2-\theta_3)
              (\theta_3-\theta_1),\\
{\cal M}_1 &=(\theta_2-\theta_3)
 \big[\theta_1^2-\theta_1(\theta_2+\theta_3)
      -2\theta_2\theta_3\big],\\
 {\cal M}_2&=(\theta_3-\theta_1)
 \big[\theta_2^2-\theta_2(\theta_3+\theta_1)
      -2\theta_3\theta_1\big],\\
 {\cal M}_3 &=(\theta_1-\theta_2)
 \big[\theta_3^2-\theta_3(\theta_1+\theta_2)
      -2\theta_1\theta_2\big]
\end{split}
\end{equation}
where again  ${\cal M}_1+{\cal M}_2+{\cal M}_3=0$. Under cyclic permutations ${\cal S}_1, {\cal S}_2$ are invariant and $\mathcal M_1\longmapsto\mathcal M_2 \longmapsto\mathcal M_3\longmapsto\mathcal M_1$. 

The symmetry classification concerns the tensor-index stripped
part of the kernel $K$. For that purpose, permutations of the
arguments must be accompanied by permutations of the indices
$I_1,\ldots,I_N$, as required by
\eqref{eq:qns3-slot-symmetry}. Relations between different ordered
regions and cyclic changes of the cut must also be imposed.

The polynomial construction is exact, but its leading-degree
argument does not exclude nonpolynomial solutions.
Section~\ref{sec:Global-kernels} will establish polynomial
exhaustion for a specified regular class of unprojected
representatives. The $\PP$ and $\PPhi$ conditions and the
physical pairing remain separate requirements.

\section{Regularity and coincidence matching conditions}
\label{sec:Global-kernels}
The construction of section~\ref{sec:Polynomial-solutions} yields polynomial 
solutions of the master equation \eqref{eq:qns3-unprojected-eigenvalue} for the tensor-index stripped kernel, within a given ordering of the insertion
positions. For a closed string, one must also ask whether such a solution
extends to a periodic tensor kernel on the full configuration space.
The extensions across neighboring ordered regions must agree, and the
kernel equation must hold without distributional sources at their common
boundaries. These regularity, continuity, smoothness, coincidence and periodicity requirements are not automatic. Here and below, coincidence refers to two insertion positions becoming equal on the
worldsheet circle, rather than to an intersection in target space.

\subsection{Joining ordered regions and regularity at region boundaries, \texorpdfstring{$N=3$}{N3} case}
\label{subsec:qns5-matching}

A kernel is periodic in each insertion position and invariant under
simultaneous permutations of position and tensor slots, cf. \eqref{eq:qns3-slot-symmetry}. The latter
symmetry relates adjacent ordered regions; it does not imply symmetry
of the tensor indices alone. 

Let us start with the $N=3$ case, for which we have the master equation \eqref{eq:qns3-cubic-equation}.
The coincident points with vanishing $\theta_i, i=1,2,3$ are pairwise coincidences on the closed null string spatial direction.  If $K$ is continuous across a smooth interface, with finite one-sided derivatives, source-free matching requires
\begin{equation}\label{eq:qns5-interface}
 (n^aA_{ab}n^b)[\partial_nK]=0,
 \qquad
 A=\begin{pmatrix}\Upsilon(\theta_1)&\theta_1\theta_2/2\\ \theta_1\theta_2/2&\Upsilon(\theta_2)\end{pmatrix}.
\end{equation}

Here $n^a$ is a unit normal 2-vector, $\partial_n=n^a\partial_a$ and brackets denote the jump. This follows
from the singular part of the Hessian, as derived in
Appendix~\ref{app:qnsE-interfaces}. Every coincidence point is characteristic: its normal second-derivative
coefficient vanishes. Finite first-derivative jumps across these points
therefore produce no source. Values must still agree, and cyclic changes
of the cut and tensor-slot exchange must be consistent. Smooth extension
from either side is not the same as equality of all derivatives across
the point. There are no characteristic hypersurfaces in the interior, as 
\begin{equation}\label{eq:b0-cubic-ellipticity}
 \det A=\frac\pi2 \theta_1\theta_2\theta_3 >0,
 \qquad A_{11}<0,\quad A_{22}<0.
\end{equation}

\paragraph{Continuity at the coincidence points.} Suppose that the kernel and its derivatives through second order
remain bounded up to $\theta_1=0$, and set $F(\theta)=K(0,\theta)$. The coefficients
of normal and mixed derivatives vanish on the coincident point, yielding
\begin{equation}\label{eq:qns5-restriction}
 \Upsilon(\theta)F''(\theta)=\frac{\mu^2}{2}F(\theta).
\end{equation}
If $\mu^2\neq0$ and $F$ has bounded second derivatives at the endpoints,
\eqref{eq:qns5-restriction} forces $F(0)=F(2\pi)=0$. A nonzero
regular $F$ then has (cf. discussions in Section~\ref{sec:quadratic-wavefunctionals}), 
\begin{equation}\label{eq:qns5-conditional-levels}
 \mu^2=\ell(\ell+1),\qquad
 F(\theta)=\left[1-\big(\frac{\theta-\pi}{\pi}\big)^2\right]
 P_\ell'\!\left(\frac{\theta-\pi}{\pi}\right),\qquad \ell\geq1.
\end{equation}
Note that no zero-average condition has been imposed on this coincidence point value and hence in itself it is
not an independently normalized quadratic kernel.
The exclusion of $\ell=1$ in \eqref{eq:b0-quadratic-levels} therefore
does not apply to this restriction.

Coincident-point values alone do not determine the interior. A nonzero
eigenfunction can vanish on every coincidence point; $\theta_1\theta_2\theta_3$ for $N=3$ case is the first
cubic example. That sector can be solved directly, rather than excluded.

\subsection{Weighted spectrum for kernels vanishing on every coincidence point, generic \texorpdfstring{$N$}{N}}
\label{subsec:qns5-dirichlet}

We are now ready to extend the analysis of previous subsection to generic $N$. It is handy to change the normalization of the separation variables $\theta_a$ and introduce another dependent-variable $x_N$:
\begin{equation}\label{eq:b0-simplex}
 x_a=\frac{\theta_a}{2\pi},\qquad x_N=1-\sum_{a=1}^{N-1}x_a,
\qquad 0<x_i<1, \qquad i=1,2,\cdots, N.
\end{equation}
The operator becomes
\begin{equation}\label{eq:b0-simplex-operator}
 L_N=-\frac12\sum_{a,b=1}^{N-1}
 (x_a\delta_{ab}-x_ax_b)\partial_a\partial_b.
\end{equation}
It is the negative of the neutral Wright--Fisher diffusion generator
without mutation. Its polynomial structure is related to the standard
multivariate Jacobi problem \cite{STEINRUCKEN20131}; we give the
needed derivation here, without importing a population-genetic interpretation.

We note that, 
\begin{equation}
 L_Nh_N=\frac12N(N-1)h_N,    \qquad h_N(x):=\prod_{i=1}^Nx_i,
\end{equation}
and that $h_N$ vanishes on all coincidence points. Writing $K=h_Nf$ gives
\begin{equation}\label{eq:b0-h-transform}
 h_N^{-1}L_N(h_Nf)=\frac{N(N-1)}2f-\mathcal J_N f,
\qquad
 \mathcal J_N=\frac12\sum_{i,j}(x_i\delta_{ij}-x_ix_j)\partial_i\partial_j
 +\sum_i(1-Nx_i)\partial_i.
\end{equation}
{
Let $\Delta_{N-1}=\{x_i>0,\ \sum_{i=1}^Nx_i=1\}$, with $dx=dx_1\cdots dx_{N-1}$. The operator $\mathcal J_N$ is symmetric in $L^2(\Delta_{N-1},h_Ndx)$. Its orthogonal polynomial eigenspaces of exact degree $j$ have eigenvalue $-j(j+2N-1)/2$ and together form a complete basis. Multiplication by $h_N$ is unitary from this space to
\begin{equation}\label{eq:b0-kernel-space}
 \mathscr H_N^{(0)}=L^2(\Delta_{N-1},h_N^{-1}dx).
\end{equation}

}The self-adjoint realization defined by this polynomial basis therefore
has a purely discrete spectrum,
\begin{equation}\label{eq:b0-dirichlet-spectrum}
 K_{j,\alpha}=h_Np_{j,\alpha},\qquad
 \mu_j^2=(j+N)(j+N-1),\qquad j=0,1,\ldots,
\end{equation}
with scalar multiplicity $\binom{j+N-2}{N-2}$. Appendix~\ref{app:qnsB-weighted}
specifies its domain and proves completeness. These functions are exactly
the polynomial eigenfunctions that vanish on every coincidence point. In the notation
of Section~\ref{sec:Polynomial-solutions}, $m=j+N$, so this sector starts
at $m=N$. In particular, $\theta_1\theta_2\theta_3$ is the lowest such cubic eigenfunction.

This norm is a norm on kernels in one ordered region. No identification
with a physical wavefunctional norm is made. Global slot symmetry,
projection, and the $\PP,\PPhi$ conditions can remove or relate these
solutions. Their scalar multiplicities are not physical state counts.

\subsection{Polynomial exhaustion for regular representatives}
\label{subsec:qns5-conditional}

The coincident-point reduction (reduction of level $N$ master equation to $N-1$ at coincidence points) and the weighted construction above together give a stronger statement than the polynomial ansatz. Assume that a component of $K$
extends smoothly to the closure of each ordered simplex and solves the
unprojected equation there. Smoothness is required separately in each
region; finite normal-derivative jumps across coincidences are allowed.
Then every nonzero positive-eigenvalue solution is polynomial in $\theta_i$ on each region, and
\begin{equation}\label{eq:b0-regular-exhaustion}
 \mu^2=m(m-1),\qquad m=2,3,\ldots,
 \qquad \deg K=m.
\end{equation}
The zero-eigenvalue solutions on one closed simplex are linear functions in $\theta_i$.
The statement holds componentwise and does not presuppose a highest
polynomial degree.

The proof is by induction on the number of positions. Restriction to
coincident points has the same eigenvalue and, by the induction hypothesis,
polynomial boundary values. Their compatible values admit a polynomial
extension into the simplex. The difference vanishes on every coincidence point and
can be divided by $h_N$. Its Jacobi expansion then contains only finitely
many degrees, because the inhomogeneous term is polynomial and the
Jacobi eigenvalues are strictly increasing. If all coincidence point values vanish,
\eqref{eq:b0-dirichlet-spectrum} applies directly. Appendix~\ref{app:qnsE-exhaustion}
supplies the extension and solvability argument, including resonant
eigenvalues and the zero-eigenvalue case.

For a periodic tensor kernel in this class, only finitely many ordered
regions occur at fixed $N$. Global matching imposes linear relations
between their finite-dimensional polynomial eigenspaces. It cannot
introduce new eigenvalues. This proves discreteness of the allowed
eigenvalues for regular representatives, including the modes invisible
to all coincidence point restrictions.

\subsection{Projection and the physical interpretation}
\label{subsec:qns5-support}

{
In the above we have crucially used the regularity hypothesis, e.g. $K=\theta_1/(\theta_1+\theta_2)$ solves the local cubic equation with zero eigenvalue but has no limit at the doubly coincident point $\theta_1=\theta_2=0$.

To obtain a nonzero state in the separated sector, a globally matched tensor kernel must have a nonzero zero-average projection satisfying \eqref{eq:qns3-Q-condition} and \eqref{eq:qns3-R-condition}. The solution ${\cal S}_1$ in \eqref{N3-m2} illustrates
why solving the ($I$-indices stripped) kernel equation alone is insufficient.
Its periodic continuation satisfies
\begin{equation}\label{eq:qns5-invisible-example}
 \widehat{\cal S}_1=-\sum_{a<b}\Upsilon(\sigma_a-\sigma_b).
\end{equation}
Each summand is independent of one insertion position,
whose integration gives
$\int_0^{2\pi}d\sigma\,\uQ^I(\sigma)=0$.
Consequently, with a constant fully symmetric tensor,
this kernel represents the zero wavefunctional.

The results establish the weighted spectrum of the coincidence point vanishing sector and polynomial exhaustion of smooth closed-region representatives. Nontrivial simultaneous solutions of the contraction conditions at higher order and their physical pairing remain to be established. The kernel norm is not a wavefunctional norm.
}
\section{Discussion and outlook}\label{sec:discussion}
{
In this work we continued the analyses of quantized null string theory on flat background started \cite{Rasulian:2026jvg}. Despite the fact that the constraint $C_3$ breaks translation symmetry on the background Minkowski space, the vacuum state \eqref{eq:VacuumCondition} retains this symmetry in transverse directions, while one still has $p^+, p^-, p^1$ on the vacuum state is nonzero ($2p^+ p^-=p_1^2$ and $p^+q^-=p_1 q_1$). That is, the vacuum state corresponds to a massless state moving in $\pm, 1$ directions. This is to be compared with the standard tensile closed bosonic string whose vacuum state is a tachyon moving probing the whole flat target spacetime. 

Our analyses also delimit the statement about dimension. The
reduced differential equations and the scalar kernel construction select
no critical target dimension. Tensor multiplicities do depend on $D$,
and a complete claim about quantum consistency would additionally require
the appropriate symmetry generators, domains, and regulator limits.

We showed within the quantization scheme discussed in detail in this paper that null strings have a discrete spectrum with the effective energy unit set by the light-cone solution space parameter $q_1$. It is instructive to explore how this fact (effective over all mass scale dovetails with the null string holography \cite{Sheikh-Jabbari:2026rcr} where strings on $D$ dimensional flat space is described by ILST theory on dS$_{D-1}$ \cite{Sheikh-Jabbari:2026rcr, Lindstrom:2026tua}. The natural guess is that in the  ``dual'' ILST formulation, $q_1$ is related to the radius of the de Sitter (dS) space. Since the duality works for dS at any radius, $q_1$ remains a free (unspecified) parameter, in accord with out analysis here.

We constructed our solutions for the kernel of the wavefunctional within physically well-motivated assumptions where the solutions appear as finite degree $m$ polynomials associated with states of mass $\mu^2=m(m+1)$. The same levels arise without assuming polynomiality for representatives
that are smooth up to the boundary of each ordered region. Coincidence point
restriction reduces the number of insertion positions, and the sector
vanishing on all coincidence points admits a complete Jacobi spectral resolution.
Together these facts give the exhaustion argument of
Section~\ref{subsec:qns5-conditional}. The proof includes modes whose
coincidence values vanish, which could not be controlled by coincidence  values
alone. It is a result about a specified class of kernel representatives.

There remain limitations in identifying these results with the spectrum
of a physical null-string Hilbert space.   The $\PP$ and
$\PPhi$ equations are conditions on a separated family rather than
additional gauge constraints. Their common solution space must be
determined at each order; invariance of that space under the composite
operator should still be established in general.

There is also a concrete issue with the physical pairing. At a finite
Fourier-mode cutoff, the standard flat $L^2$ representation makes
$\PP^{II}=\sum_{n\neq0}(P_n^I)^\dagger P_n^I$ nonnegative. A vector
in its operator domain with $\PP^{II}\psi=0$ is annihilated by every
$P_n^I$. On noncompact Fourier coefficient space there are no nonzero
square-integrable vectors with this property for every $I$. The formal
separated polynomial states therefore cannot simply be declared
normalizable in that representation. A justified generalized pairing,
or a different physical representation with its adjoints and domains
specified, is needed. The weighted kernel norm in
\eqref{eq:b0-kernel-space} does not resolve this issue.

Finally, so far we have analyzed free minimal  null strings on flat backgrounds. The next steps involve moving away from minimality, e.g. along the lines discussed in \cite{Sheikh-Jabbari:2026rcr, Sheikh-Jabbari:2026nez} and introducing null string interactions. For the latter one should classify CW-invariant vertex operators. 

}\color{black}


\paragraph{Acknowledgments.}
MMShJ acknowledges Iranian National Science Foundation (INSF) research chair grant No.~4041653.  

\appendix

\section{Reduction and composite ordering}
\label{app:qnsA-reduction}
\label{app:qnsB-operators}
The light-cone reduction applies on a patch where $P^+$ and the
chosen component $Q^1$ do not vanish. The remaining rigid
$\sigma$-translation imposes $G_0=0$. At fixed $p^+$, the
symplectic potential, up to an exact variation, is
\begin{equation}\label{eq:b0-reduced-one-form}
 \Theta_{\mathrm{LC}}
 =p_1\delta q_1+p^I\delta q^I
 +\sum_{n\neq0}P_{-n}^I\delta Q_n^I.
\end{equation}
Thus $(q_1,p_1)$ is a global canonical pair. Holding both fixed
defines the conditional transverse quantization used here.

\paragraph{Composite prescription.}
\label{app:qnsB-product}
Composing the operators in \eqref{eq:b0-product-prescription}
gives the second-order term in \eqref{PhiPhi-quantum} and a
first-order coefficient
\begin{equation}\label{eq:b0-contraction-sum}
 \sum_{\substack{0<|n|\leq L\\n\neq-r}}
 \frac{r(r+n)}{n^2}
 =2r^2\sum_{n=1}^L\frac1{n^2}
 \longrightarrow\frac{\pi^2r^2}{3}.
\end{equation}
The excluded numerator vanishes, and the terms proportional
to $r/n$ cancel. No scalar term appears because every
$\widehat\Phi_n$ annihilates constants. This fixes $b=0$
within the prescribed composite product.

A separate cutoff on the Fourier coefficients can change
finite-cutoff edge terms. The coefficientwise prescription
used here removes that cutoff first, at fixed transfer cutoff
$L$, and subsequently takes $L\to\infty$. This calculation
does not establish convergence in a physical operator norm.
\section{Ordered regions and polynomial completeness}
\label{app:qns-polynomial-recursion}
\label{app:qns-matching}
\paragraph{Ordered-region operator.}
\label{app:qnsD-chamber}
\label{app:qnsD-scope}
For the increasing ordering in \eqref{eq:qns4-gaps},
\begin{equation}\label{eq:qnsD-derivatives}
 \partial_{\sigma_i}
 =\partial_{\theta_{i-1}}-\partial_{\theta_i},
 \qquad
 \partial_{\theta_0}=\partial_{\theta_N}=0.
\end{equation}
Set $s_{ij}=\sigma_j-\sigma_i
=\sum_{a=i}^{j-1}\theta_a>0$ for $i<j$.
Translation invariance and summation across each gap give
\begin{align}
 \sum_{i<j}s_{ij}^2\partial_{\sigma_i}\partial_{\sigma_j}K
 &=-\mathcal D(\mathcal D-1)K,\\
 \sum_{i<j}s_{ij}\partial_{\sigma_i}\partial_{\sigma_j}K
 &=-\sum_a\theta_a\partial_{\theta_a}^2K.
\end{align}
Substitution in $L_N=-\mathscr L$ proves
\eqref{eq:qns4-operator}. The polynomial recursion and its
solution are already given in the main text.

\paragraph{Weighted completeness.}
\label{app:qnsB-weighted}
Put $a_{ij}=x_i\delta_{ij}-x_ix_j$, with
$i,j=1,\ldots,N-1$. This matrix is positive definite in the
interior by weighted Cauchy--Schwarz, since
$\sum_{i=1}^{N-1}x_i<1$. Differentiation gives
\eqref{eq:b0-h-transform} and
\begin{equation}\label{eq:b0-jacobi-divergence}
 \mathcal J_Nf
 =\frac1{2h_N}\partial_i(h_Na_{ij}\partial_jf).
\end{equation}
Boundary flux vanishes for functions smooth on the closed
simplex, so $\mathcal J_N$ is symmetric in $L^2(h_Ndx)$.

Let $\mathscr W_j$ be the orthogonal complement of polynomials
of degree at most $j-1$ within those of degree at most $j$.
The operator preserves polynomial degree and acts on the
highest-degree terms by $-j(j+2N-1)/2$. Symmetry then implies
that every element of $\mathscr W_j$ has this eigenvalue.
Polynomial density in the finite measure $h_Ndx$ proves
completeness, and
$\dim\mathscr W_j=\binom{j+N-2}{N-2}$.

Multiplication by $h_N$ is unitary to
$\mathscr H_N^{(0)}$. The resulting diagonal operator is
essentially self-adjoint on the finite eigenbasis span.
Its closure has domain
\begin{equation}\label{eq:b0-weighted-domain}
 \left\{
 K=\sum_{j,\alpha}c_{j,\alpha}h_Np_{j,\alpha}:
 \sum_{j,\alpha}
 \left(\frac{\mu_j^2}{2}\right)^2|c_{j,\alpha}|^2<\infty
 \right\},
\end{equation}
with $\mu_j^2$ given in \eqref{eq:b0-dirichlet-spectrum}.
Finite multiplicities and eigenvalues tending to infinity
give compact resolvent.

\paragraph{Matching and coincident point restriction, derivation of \eqref{eq:qns5-interface}.}
\label{app:qnsE-interfaces}

For a continuous piecewise smooth kernel, the singular Hessian
at an interface is
\begin{equation}
 (\partial_i\partial_jK)_{\mathrm{sing}}
 =n_in_j[\partial_nK]\delta_\Sigma,
\end{equation}
where $n_i$ is a unit normal vector, $\partial_n=n^i\partial_i$ and brackets denote the jump. The normal coefficient vanishes on every coincidence point,
allowing finite derivative jumps there. The interior is
elliptic. Bounded derivatives through second order permit
restriction to a coincidence point, giving the same eigenvalue equation
for $L_{N-1}$; cyclic relabeling treats the final coincidence point.
Tensor slots are unchanged.

\paragraph{Polynomial exhaustion.}
\label{app:qnsE-exhaustion}
Suppose $K$ extends smoothly to the closed simplex and
$L_NK=\lambda K$. We prove polynomiality by induction on $N$.
On an interval, $\lambda\neq0$ forces zero endpoint values,
and the weighted resolution above applies. At $\lambda=0$,
the solutions are affine.

If $K$ vanishes at every restriction to a coincident point, successive smooth division
by the coincidence point defining functions gives $K=h_Nf$, with $f$ smooth.
The Jacobi resolution makes $f$ polynomial and fixes
$\lambda=(j+N)(j+N-1)/2$.

Otherwise, for $\lambda\neq0$, induction gives
$\lambda=m(m-1)/2$ and compatible polynomial coincidence point values
of degree at most $m$. Barycentric interpolation extends
them to a polynomial ${\cal P}$ of degree at most $m$.
Write $K-{\cal P}=h_Nf$. Restriction to coincident points implies
\[
 (L_N-\lambda){\cal P}=h_N {\cal Q},\qquad \deg {\cal Q}\leq m-N,
\]
where ${\cal Q}=0$ if $m<N$. Hence
\begin{equation}\label{eq:b0-inhomogeneous-jacobi}
 \left[\frac{N(N-1)}2-\mathcal J_N-\lambda\right]f=-{\cal Q}.
\end{equation}
On $\mathscr W_j$, the coefficient on the left is
\[
 \frac{(j+N)(j+N-1)-m(m-1)}2.
\]
For $m<N$ it never vanishes, giving $f=0$.
For $m\geq N$, it vanishes only at $j=m-N$.
Existence of the assumed smooth solution requires the
corresponding component of ${\cal Q}$ to vanish. All components
of $f$ above degree $m-N$ vanish, so $K$ is polynomial;
its highest degree is $m$ by \eqref{eq:qns4-operator}.

For $\lambda=0$, the affine coincidence-point values have one common
affine extension ${\cal P}$, determined by the vertices.
The difference $K-{\cal P}$ vanishes on every coincident point and cannot
have zero eigenvalue in the positive weighted sector.
Thus $K={\cal P}$. Smoothness and the vanishing boundary flux
justify all Jacobi coefficient equations used above.

\providecommand{\href}[2]{#2}\begingroup\raggedright\endgroup


\end{document}